\documentstyle[psfig,aps,prl,epsf,floats]{revtex}
\begin{document}
\twocolumn[\hsize\textwidth\columnwidth\hsize\csname %
@twocolumnfalse\endcsname
\title{Effect of magnetic fluctuations on the normal state properties of Sr$_2$RuO$_4$}
\author{Canio Noce, Gaetano Busiello, and  Mario Cuoco }
\address{I.N.F.M. -Unit\`a di Salerno\\ Dipartimento di Scienze
Fisiche ''E.R. Caianiello'', \\ Universit\`a di Salerno \\ I-84081
Baronissi (Salerno), ITALY}
\maketitle
\begin{abstract}
We investigate the normal state transport properties of
Sr$_2$RuO$_4 $ and we show that a consistent explanation of the
experimental results can be obtained assuming that the system is
near a quantum phase transition. Within the framework of a
self-consistent spin fluctuation theory, we calculate the
temperature variation of some relevant physical quantities and we
discuss a possible microscopic origin of the quantum phase
transition.
\end{abstract}
]
\narrowtext
\bigskip
Since the discovery of superconductivity in the layered
perovskite oxide Sr$_2$RuO$_4$ $\cite{maeno}$ much attention has
been paid to the electronic, magnetic and superconducting
properties of this compound. In its stochiometric composition
Sr$_2$RuO$_4$ is metallic, shows a Fermi liquid behavior below a
certain temperature, undergoes to a superconducting transition at
the rather low transition temperature of $\sim $1K.

Recent experiments lead to speculate that the electrons in the
Cooper pairs would have aligned spins indicating a triplet
superconductivity, and an odd orbital wavefunction as for example
p-wave \cite{andy2}-\cite{rise}, in contrast with the spin-singlet
pairing of conventional superconductors as well as of the
high-$T_c$ copper oxides.

Its electronic properties are mainly determined by three bands
crossing the Fermi level produced by antibonding Ru 4d$\epsilon $
(xy,yz,xz) and O 2p$\pi $ orbitals $\cite{mac}$.

It is found that the electrons belonging to the xy band play an
important role determining the normal state properties. From
$^{17}$O NMR in $^{17}$O isotope-enriched measurements
$\cite{imai},\cite{muku} $, spin correlations in Ru d$_{xy}$ and
d$_{xz,yz}$ orbitals are probed separately. They provide the
evidence that only the spin susceptibility $\chi _{xy}$ from the
electrons in the d$_{xy}$ orbital shows significant temperature
dependence. $\chi _{xy}$ increases monotonically with decreasing
temperature down to about 40K, turns over, and then tends to level
below T$_{FL}\equiv $25K in the canonical Fermi liquid state,
implying that the spin correlations are predominantly
ferromagnetic in origin, and orbital dependent. We point out that,
below T$_{FL}$ the out-of-plane and in-plane resistivity have a
quadratic temperature dependence indicating that a crossover to a
Fermi-liquid state takes place below 25K prior to the
superconducting transition. On the other hand, above T$_{FL}$ the
in-plane resistivity show a superlinear behaviour which may be
related to scattering due to spin fluctuations. These experimental
results indicate a clear connection to the crossover in the
magnetic transport measurements.

The demonstration that spin correlations in the RuO$_2$ planes are
predominantly ferromagnetic comes also from measurements of
nuclear spin-lattice relaxation rate.\cite{imai} The hyperfine
form factor for the $^{101}1/T_1 T$ at the Ru site is costant in
the reciprocal space, while is proportional to $\cos(q_{x,y}/2)^2$
for $^{17}1/T_1 T$ at the planar oxygen O site. This means that
the $^{101}1/T_1 T$ can probe both ferromagnetic and
antiferromagnetic spin fluctuations. On the other hand, since the
form factor for the planar O site is zero at the staggered wave
vector ${\bf q}=\{\pi,\pi\}$, antiferromagnetic spin fluctuations
do not contribute to $^{17}1/T_1 T$ \cite{thu97}. Hence, one can
probe whether the spin correlations are ferromagnetic or
antiferromagnetic by comparing the $^{101}1/T_1 T$ and $^{17}1/T_1
T$. It turns out\cite{imai} that both $^{101}1/T_1 T$ and
$^{17}1/T_1 T$ increase monotonically down to $T_{FL}$, and almost
saturate in a Korringa-like behaviour.

Besides, inelastic neutron scattering measurements\cite{sid99} in
the normal state reveal the existence of incommensurate magnetic
spin fluctuations located at ${\bf q}=(\pm 0.6 \pi /a,\pm 0.6 \pi
/a,0)$ due to the pronounced nesting properties of the almost
one-dimensional $d_{xz,yz}$ bands. Infact, the 1D sheets can be
schematically described by parallel planes separated  by
$\check{q}=\pm 2 \pi /3a$, running both in the $x$ and in the $y$
directions which give rise to dynamical nesting effects at the
wave vectors ${\bf k}=(\check{q},k_y)$, ${\bf k}=(k_x,\check{q})$,
and in particular at ${\bf
\check{q}}=(\check{q},\check{q})$.\cite{prl99}

Combining the results of NMR,\cite{imai} nuclear spin-lattice
relaxation rate,\cite{imai} and inelastic neutron scattering
measurements\cite{sid99}, it is possible to draw the following
physical picture for the Sr$_2$RuO$_4$. There is a strong
enhancement of ferromagnetic spin correlations above T$_{FL}$
mainly due to ferromagnetic correlations between the electrons in
the $d_{xy}$ band as revealed from Knight shifts experiments.
These contributions are in competition with incommensurate spin
fluctuations due to the nesting properties of the almost 1D
$d_{xz,yz}$ bands. Hence, the main features of the magnetic
response turn out to be decoupled for the electrons in the
$d_{xy}$ and in the $d_{xz,yz}$ bands, respectively.

The aim of this Letter is to propose an explanation of the non
Fermi liquid behaviour observed in the $\chi_{xy}$ and in the
$1/T_1 T$ experiments of the Sr$_2$RuO$_4$ assuming that this
compound is close to a 2D ferromagnetic quantum critical point.
Then, we will discuss how it might be possible to drive the system
from a magnetic to a non-magnetic state considering the interplay
between the lowering of the crystal symmetry and the magnetic
correlations.
\begin{figure}
\centerline{\psfig{figure=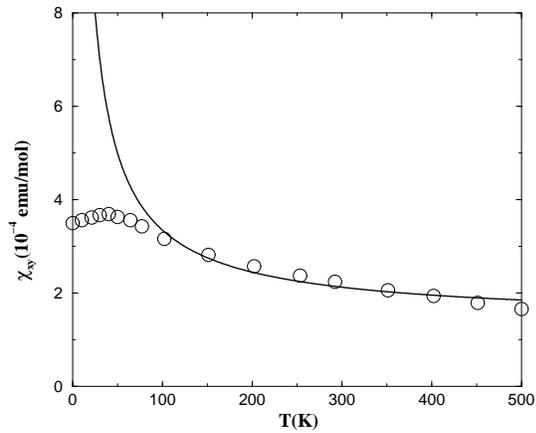,width=7cm}} \caption{Magnetic
susceptibility for the d$_{xy}$ orbital. The circles stand for the
experimental results from $\protect\cite{imai}$ and the solid line
is the theoretical curve.} \label{fig1}
\end{figure}
Let consider the effect of ferromagnetic fluctuations in the
correlated $d_{xy}$ band considering the system as a 2D
ferromagnet with vanishing critical temperature. The starting
point in our case is represented by the correlated dynamics in the
RuO$_2$ planes of the $d_{xy}$ electrons, which we consider
responsible for ferromagnetic fluctuations as suggested from NMR
and nuclear spin-lattice relaxation rate.

In this case as in many others metallic systems at low
temperatures $\cite{lonz1}$, it has been observed the signature
for a non-Fermi-liquid behavior near a quantum phase transition.
The proximity to a quantum critical point is usually marked by
enhancement in the effective mass, and in spin and charge response
at low temperatures. This in turn introduces a low energy scale
which marks a crossover from quantum to a classical behavior in
the temperature dependence of various physical properties.

Now, to calculate some relevant physical properties, we consider
the self-consistent spin fluctuation theory developed by
Ramakrishnan {\it et al} $\cite{rama}$ and many others
$\cite{mori}$ for the $d_{xy}$ band.

The main motivation for the use of spin-fluctuation theory is the
largeness of the Stoner enhancement factor $1/\alpha(0)$ at $T=0$.
In such a case, a highly paramagnetic system at low temperature
can be considered to be in the vicinity of a magnetic transition.
The variation of relevant physical quantities is therefore
governed by transverse and longitudinal spin fluctuations. Even
though the order parameter is zero above the transition, the
effect of fluctuations is observable well above the transition.

From a general point of view, the susceptibility can be obtained
from the equation of state for the external magnetic field $H$ and
the magnetization $M$
\begin{eqnarray}
\frac{H}{M}=\alpha(T)+\beta M^2 \label{alfaMF}
\end{eqnarray}
where the coefficient $\alpha$ and $\beta$ took different
expressions depending on the approximation. In the mean field
theory of itinerant ferromagnets $\alpha_{MF}(T)=1-U
\rho(\epsilon_F)$ and $\beta$ is constant, $ \rho(\epsilon_F)$
being the density of states at the Fermi level. As the temperature
dependence is weak, of the order of $(T/T_F)^2$, one cannot yield
a Curie-Weiss form of the spin-susceptibility. Here T$_F$ is the
Fermi temperature. This issue is overcome in the spin fluctuation
theory where $\alpha(T)$ is given by \cite{rama,mori}:
\begin{eqnarray}
\alpha(T)=\alpha(0)+u(2D^T+3D^L),
\end{eqnarray}
here $\alpha(0)$ is the susceptibility enhancement factor at
$T=0$, which includes the mean field part of Eq. \ref{alfaMF} and
the zero-temperature part of the fluctuation self-energy. Still,
$D^T$ and $D^L$ are the transverse and longitudinal spin
fluctuation amplitudes obtained by a diagrammatic expansion of the
effective Hamiltonian, after integrating over the electronic
degrees of freedom. The factor $u$ is a dimensionless short range
fluctuation coupling constant\cite{rama}. In this framework, the
susceptibility for the ferromagnetic case has the following form:
\begin{eqnarray}
\chi({\bf q},\omega)=\frac{\rho(\epsilon_F)}{\alpha(T)+\delta {\bf
q}^2-i \frac{\pi \omega \gamma}{2 {\bf q}}} ,
\end{eqnarray}
where the parameters $\delta$ and $\gamma$ depend on the
diagrammatic expansion.

It is found that there are two different regimes which for a 2D
ferromagnet read as follows: when $\tau <\alpha (0)$, one has an
enhanced Pauli susceptibility behaviour, while at higher
temperatures, i.e. for $\alpha (0)<\tau <1,$ the susceptibility
follows the classical Curie Weiss behavior. Here, $\tau $=T/T$_F$.
Therefore, the quantity $\alpha (0)$T$_F$ plays the role of the
low energy scale and naturally comes out from the fluctuation
theory. The comparison between the experimental data from Imai
{\it et al} $\cite{imai}$ and the xy spin susceptibility
calculated within the above assumptions, is reported in
Fig.\ref{fig1}.
\begin{figure}
\centerline{\psfig{figure=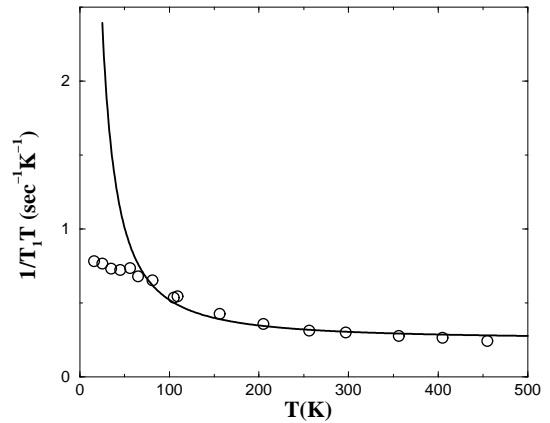,width=7cm}}
\caption{Nuclear spin-lattice relaxation rate at Ru site vs
temperature: the solid line stands for the theoretical result and
the circles are the experimental data taken from
$\protect\cite{imai}$.} \label{fig2}
\end{figure}
As we can see, there is a good agreement between the experimental
results and the theoretical prediction. It is worth stressing that
using the experimental data for the spin magnetic susceptibility
and the Fermi temperature, calculated using the Fermi wavevector
and the effective mass reported by Mackenzie {\it et al } $\cite
{mac}$, we obtain for $\alpha (0)$T$_F$ the value $\sim $70K which
gives the right order of magnitude for the temperature T$_{FL}$ of
the crossover into a Fermi liquid state.

Within the same approach it is easy to obtain the temperature
dependence of the nuclear spin-lattice relaxation rate.

One gets $(T_1 T)^{-1} \sim \alpha(T)^{3/2}$ $\cite{rama}$; for a
normal Fermi-liquid $\alpha(T)$ is constant and gives the usual
Korringa law, while for a 2D ferromagnet varies as $T^{-1}$.

Assuming the previous temperature dependence for (T$_1$T)$^{-1}$
for the xy electrons and a usual Korringa law for the other two
bands, we obtain a fair fit, shown in Fig.\ref{fig2}, to the
experimental results of Imai {\it et al} $\cite{imai}$ for
temperatures larger than T$_{FL}$.

It is worth pointing out that though the contributions of the
$d_{xz,yz}$ contain the effect of dynamical nesting, in the
present calculation they have been added at a mean field level so
that there is a weak temperature dependence of the order of $\sim
(T/T_F)^2$ in the range examined. If we consider the correction
due to spin fluctuations instead that at ${\bf q}=0$ around the
nesting vector ${\bf \check{q}}=(\check{q},\check{q})$ we would
expect a contribution to (T$_1$T)$^{-1}$ with a temperature
dependence $\sim T^{-1}$. A more detailed analysis considering the
multi-orbital system within the spin fluctuation theory is in
progress.
\begin{figure}
\centerline{\psfig{figure=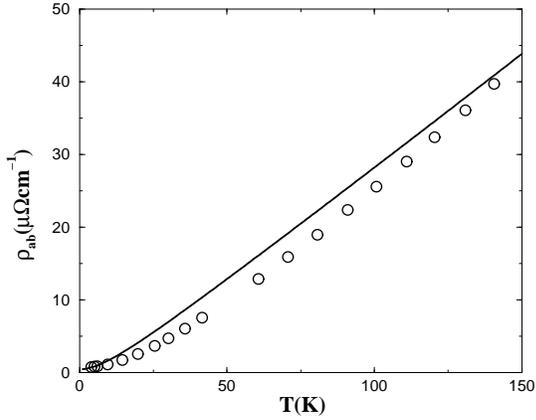,width=7cm}}
\caption{In plane resistivity vs temperature: the circles are the
experimental results from  $\protect\cite{nig}$ while the solid
line is the theoretical curve.} \label{fig3}
\end{figure}
Furthermore, we expect that the transport properties are affected
by spin-fluctuation above T$_{FL}$ and that this effect should
manifest mainly in the in-plane quantities. Therefore, we have
calculated the in-plane resistivity using the Boltzmann equation
for a multi-band anisotropic electronic system assuming for the
relaxation rates, two Fermi liquid ($\tau_{f}^{-1}\sim T^2$)
behaviors and a two-dimensional ferromagnetic decaying time
$\tau_{sf}^{-1} \sim T$ as due to the scattering of the spin
fluctuations with the electrons in the $d_{xy}$
band\cite{zim,mathon}.

It is worth pointing out that the relaxation rates $\tau_{f}$ have
been determined in a way to reproduce the in-plane Hall
coefficient and the magnetoresistance, respectively \cite{CM}. The
fit to the experimental data reported in Fig.\ref{fig3}, where the
experimental data are taken from $\cite{nig}$, indicates that the
assumption of linear temperature dependence in the decaying
relaxation rate of the xy electron-dynamic is essentially correct.

Let consider now the origin of the ferromagnetic spin fluctuations
looking at the problem from a more general point of view.

One argument on why the ferromagnetic spin fluctuations play an
important role in determining the transport properties of
Sr$_2$RuO$_4$ is the following. The Ruddlesen-Popper series
Sr$_{n+1}$Ru$_n$O$_{3n+1}$ contains multilayer compounds
with $n$ as the number of RuO$_2$%
-planes per unit cell. In particular, there is a series of
ferromagnetic compound related to Sr$_2$RuO$_4$: the
infinite-layer (3D) SrRuO$_3$ is ferromagnetic with T$_c\simeq
$165K $\cite{srruo3}$; for n=3 one finds T$_c\simeq $148K
$\cite{n3}$ and for n=2 T$_c\simeq $104K $\cite{n2}$, although the
ferromagnetism in this last case is still controversial. These
values indicate the tendency that with decreasing the layer number
n, T$_c$ is reduced and finally vanishes\cite{rice}, since for
$n=1$ the system is paramagnetic at all temperatures before
becoming superconductor below $\sim 1 K$.

Since $n$ is varying in a discrete way, it is meaningless to
consider it as the direct parameter whose continuous change
controls the transition from a paramagnetic to a ferromagnetic
phase. On the other hand, it is possible to image stacks of
multilayer Sr$_{n+1}$Ru$_n$O$_{3n+1}$ compounds with different $n$
so that the final system has an average number of planes which may
be varied continuously. In this framework, the average number of
RuO$_2$ planes can represent the parameter controlling a quantum
phase transition. This kind of ideal system can in principle be
realized by performing a molecular beam epitaxial growth.

An important point on the changes induced by $n$ is that when we
move from the cubic SrRuO$_3$ compound to the tetragonal
Sr$_2$RuO$_4$, the modification of the magnetic ground state is
accompanied by the lowering of the lattice symmetry.

We know other examples within the metal transition oxides whose
magnetic properties are strictly related to structural
changes.\cite{Kug82} These cases are mainly insulators systems
where the Jahn-Teller effect cause structural phase transitions,
and gives rise to orbital ordering and new magnetic structure.

In our situation, the removal of the degeneracy due to crystal
field effects without Jahn-Teller distortions produces a subtle
change in the magnetic properties. Due to the peculiar filling of
the $4d$ level in the SrRuO systems, we will show that the ground
state switches from a paramagnetic to a ferromagnetic
configuration depending on the way one lowers the lattice
symmetry, i.e. by elongating or compressing the RuO$_6$ octaedra
along the c-axis.

This effect might be experimentally fulfilled by applying an
external uniaxial pressure and/or by means of substitution of the
Sr$^{+2}$ with a larger ion with the same valence. In this way, it
might be possible to induce changes in the local crystal field
driving the system from a magnetic state to a non-magnetic one or
vice versa.

To have a quantitative insight on these considerations, we study
the effect of the removal of the cubic symmetry for a $d^2$ system
(two holes in the $t_{2g}$ orbitals).

The Hamiltonian is given by:
\begin{eqnarray*}
H&=&\sum_{i,\alpha} \epsilon_{i,\alpha} n_{i,\alpha} \\&+&(U+2
J_H) \sum_{i \alpha} n_{i,\alpha \uparrow} n_{i,\alpha \downarrow}
+(U-\frac{1}{2}J_H) \sum_{i,\alpha < \beta} n_{i,\alpha}
n_{i,\beta}\\ \nonumber &-&2 J_{H} \sum_{i, \alpha < \beta}
{\bf{S_{i,\alpha}}} {\bf{S_{i,\beta}}}+J_H \sum_{i,\alpha \beta}
d_{i \alpha \uparrow }^{\dagger} d_{i \alpha \downarrow
}^{\dagger} d_{i \beta \downarrow } d_{i \beta \uparrow }\\&+&
\sum_{ <i,j>\alpha \sigma} t_{\alpha}
(d_{\alpha,i\sigma}^{\dagger} d_{\alpha,j \sigma }+{\it H.c.})
\end{eqnarray*}

\noindent where the Hamiltonian contains intrasite terms, and the
diagonal hopping between electrons in the different $t_{2g}$
orbitals.

\noindent The operator $d_{\alpha,i\sigma}$ destroys an electron
with spin $\sigma$ in the orbital $\alpha$ on site $i$,
$n_{i,\alpha \sigma}$ is the electron density with spin $\sigma$
on the orbital $\alpha$, and ${\bf{S_{i,\alpha}}}$ is the spin of
the electron in the $\alpha$ orbital on site $i$, respectively.
Here, $\epsilon_{i,\alpha}$ denotes the on-site energy of the
$\alpha$ t$_{2g}$ orbital on the site $i$, $U$ and $J_H$ stands
for the Coulomb and Hund's exchange interaction, respectively, and
$t$ denotes the hopping amplitude.

The cubic-tetragonal symmetry crossover is controlled by means of
the zero energy splitting
$\Delta\equiv(\epsilon(d_{xy})-\epsilon(d_{\gamma z}))$ between
the $d_{xy}$ and $d_{xz,yz}$ orbitals.

In particular, the variations of $\Delta$ describe the interplay
between ionic and covalent effects induced by the oxygens
surrounding the Ru ion and thus simulate the effect of compressing
and/or elongating the RuO$_6$ octaedra along the c-axis.

\begin{figure}
\centerline{\psfig{figure=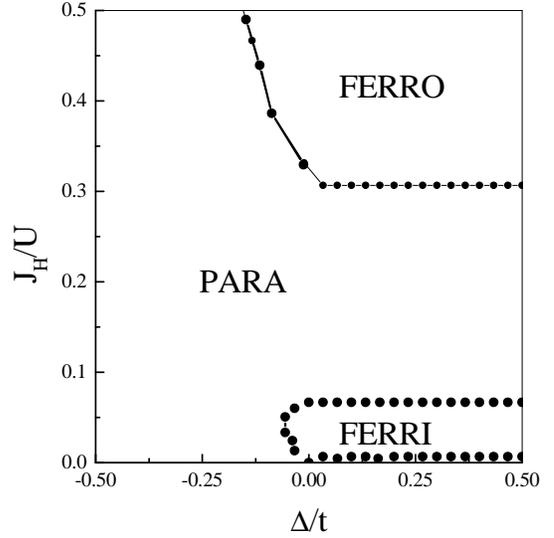,width=7cm}}
\caption{Phase diagram for the two-sites cluster. Ferro, ferri and
para stand for ferromagnetic, ferrimagnetic and paramagnetic
ground state} \label{fig4}
\end{figure}

The phase diagram is obtained considering two effective ruthenium
atoms. The results of this numerical simulation are reported in
Fig.\ref{fig4} with $J_H/U$ vs $\Delta/t$.

For the cubic case, i.e. when $\Delta$=0, the ground state is
ferrimagnetic (total spin is equal to one) in a very small region,
above that it becomes ferromagnetic passing through a spin zero
(paramagnetic) configuration. When the value of the exchange
interaction $J_H$ is large enough, crystal field effects,
associated with the elongation of the RuO$_6$ octaedra
($\Delta>$0), tend to stabilize ferromagnetic spin configurations.

Indeed, for positive $\Delta$, the $d_{xy}$ orbital is lower in
energy than the others two degenerate $d_{xz,yz}$. Then, if the
Coulomb repulsion is larger respect to $\Delta$ and the other
parameters involved, one would occupy the first lower energy
orbital and put the other hole in one of the two degenerate
$d_{xz,yz}$. In this case there is a gain in the kinetic energy if
the holes on the neighbour site have the same spin of the hole
moving. We have a kind of dynamical {\it double-exchange}
mechanism since the spin of the neighbour site is fluctuating in
amplitude and phase, instead of having only phase fluctuation as
happens in the manganite oxides. Still, strictly speaking, the
exchange mechanism does not occur between localized spins.

On the other hand, the gain in kinetic energy due to the large
$d-p$ hybridization, inverting the order of the molecular orbitals
respect to the result of crystal field\cite{canmar} and/or the
compression of the RuO$_6$ octaedra, would produce negative value
for $\Delta$ and stabilize a paramagnetic configuration.

Indeed, for negative $\Delta$, the lowest energy configuration is
realized by occupying with two holes the $d_{xz,yz}$ orbitals. In
this case for large Coulomb repulsion we would have and effective
Heisenberg antiferromagnetic Hamiltonian with a net exchange $J
\sim 4 t^2/U$ which sets the lowest spin configuration in the
ground state.

These two limiting cases come in competition when $|\Delta|$ is
small and of the order of $t$. In this case the gain in the
kinetic energy allow to have configurations with non integer
occupation number on each $t_{2g}$ orbital. This is the relevant
situation which occurs in the ruthenium oxides either for the
cubic and the tetragonal compound.
Some comments on the limits of this cluster calculation are
worthwhile. Within this two-sites problem one could not include
any band effect which, for example, is relevant to understanding
the dynamical nesting at incommensurate wave vector in the
Sr$_2$RuO$_4$. Neverthless, the present calculation yields useful
insights on the interplay between the magnetic correlations and
the orbital degrees of freedom in the complete series of the SrRuO
compounds. We point out that the study of the electron dynamics in
a $d^2$ configuration have been attracted new interest also in
other compounds like the V$_2$O$_3$, suggesting more detailed
analysis in this direction.

In conclusion, in this paper we described the non-Fermi liquid
behaviour observed in the normal state properties of
Sr$_2$RuO$_4$. We have shown that, assuming that this compound
lies very near to a quantum critical point, the temperature
variation of various physical quantities is governed by spin
fluctuations, whose effect is observed over a wide temperature
range. A consistent explanation of the experimental results is
obtained within this picture and assuming that only the electrons
belonging to $d_{xy}$ band are involved in this mechanism.
Moreover, we have given indication how to drive the system from a
non-magnetic to a magnetic configuration both on the theoretical
and experimental side which follows directly from the peculiar
interplay between the lowering of the crystal symmetry, which
involves the orbital degrees of freedom, and the spin
correlations.

%

%
\end{document}